\documentstyle[prd,preprint,aps]{revtex}

\begin{document}

\reversemarginpar

\title{The quantum absolute phase observable}
\author{Gilad Gour\thanks{E-mail:~gour@cc.huji.ac.il}
\thanks{Tel:~972-2-6586547 , Fax:~972-2-5611519}} 
\address{Racah Institute of Physics, Hebrew University of Jerusalem,\\ 
Givat Ram, Jerusalem~91904, ISRAEL.}

\maketitle

\begin{abstract}

Defining the observable ${\bf \phi}$ canonically conjugate to the number
observable ${\bf N}$ has long been an open problem in quantum theory. Here 
we show how to define the absolute phase observable 
${\bf \Phi}\equiv |{\bf\phi}|$ by suitably restricting the Hilbert space 
of $x$ and $p$ like variables.  This ${\bf \Phi }$ is actually the absolute 
value of the phase and has the correct classical limit. A correction to the 
``cosine'' ${\bf C}$ and ``sine'' ${\bf S}$ operators of Carruthers and
Nieto is obtained.

\end{abstract}
$\;\;\;\;\;\;\;\;\;\;\;\;\;\;\;$PACS numbers:~42.50.Dv,~03.65.Vf,~03.65.Ca,~03.65.Ta


\reversemarginpar

In 1927, Dirac~\cite{Dirac} proposed a phase observable $\phi$, supposedly 
a canonical conjugate to the number operator ${\bf N}={\bf a}^{\dag}{\bf a}$. 
The idea was to decompose the annihilation operator in the following 
form: ${\bf a}=\exp(-i{\bf \phi}){\bf N}^{1/2}$. Many years later, Susskind 
and Glogower~\cite{Susskind} pointed out that such a procedure was not correct
because ${\bf E}^{\dag}$, defined by  
${\bf a}={\bf E}^{\dag}{\bf N}^{1/2}$,
is not a unitary operator, and thus cannot be expressed as $\exp(-i{\bf \phi})$.
Despite the nonunitary nature of ${\bf E}$ and ${\bf E}^{\dag}$, Carruthers
and Nieto~\cite{CandN} have defined Hermitian ``sine'' and ``cosine'' operators 
${\bf S}=(1/2i)({\bf E}-{\bf E}^{\dag})$ and 
${\bf C}=(1/2)({\bf E}+{\bf E}^{\dag})$, respectively. However, these operators
cannot represents the exact sine and cosine of the phase observable
since, for example, ${\bf S}^{2}+{\bf C}^{2}\ne 1$.
Since that time many new techniques have been developed to define a 
phase operator~\cite{Carruthers,Lynch}. 
In a number of theories, such as Susskind and Glogower's~\cite{Susskind}, 
the Barnett and Pegg formalism~\cite{Pegg}, etc, 
the Hermitian phase operator is not well defined. Some other 
theories do not pass the Barnett and Pegg ``acid-test''~\cite{Barnett}:
the eigenstates of the number operator do not represent
states of indeterminate phase. In an excellent critical review, 
Lynch~\cite{Lynch} argues that there is as yet no satisfactory solution to 
the quantum phase problem. However, there is a solution if one generalize
the formal description of measurement to include the so-called POM or POVM 
observables~\cite{Shapiro}. 

In the present work we devise a new approach which succeeds in giving the 
absolute value ${\bf\Phi}\equiv |{\bf\phi}|$ of the phase observable.
As Moshinsky and Seligman~\cite{MoSe} have shown many years ago, the classical
canonical transformations to action and angle variables for harmonic oscillator
(and some other systems) turn out to be nonbijective (not one-to-one onto).
Hence, the fact that it is possible to construct only the $absolute$ phase observable,
and not the phase observable itself, is not surprising even from the classical point 
of view. 

Two problems arise in the construction of a phase observable. 
The first one (also the easier to overcome), is that
the phase operator ${\bf\phi}$ is supposed to be an angle operator. Hence, it is
restricted to a finite interval which is chosen, somewhat arbitrarily, 
to be $(-\pi ,\pi]$. Thus, the matrix elements of $[{\bf N},{\bf\phi}]$ 
in the number state basis $|n\rangle$,
\begin{equation}
\langle n|[{\bf N},{\bf \phi}]|n'\rangle=(n-n')\langle n|{\bf \phi}|n'\rangle
\end{equation}
vanish for $n=n'$ because $|\langle n|{\bf\phi}|n\rangle|\leq\pi$.
This implies that $[{\bf N},{\bf\phi}]\neq i$ so that ${\bf\phi}$ is not canonical
to ${\bf N}$. The same problem appears also in the commutation relation of an 
angular momentum ${\bf J}_{z}$
component and its associated angle ${\bf\Theta}$~\cite{Judge,BandP}. The solution
to this last problem is well known: the commutation relation 
is to be changed to 
\begin{equation}
[{\bf J}_{z},{\bf \Theta}]=i\hbar\left(1-2\pi\delta({\bf \Theta}-\pi)\right),\;\;\;\;
-\pi<\theta\leq\pi,
\label{npcr}
\end{equation}
where ${\bf\Theta}$ can be expressed as a $2\pi$-periodic function of an 
unrestricted angle operator~\cite{Lynch}.

The second problem arises due to the fact that the number operator is bounded 
from below. As we now show, the commutation relation like~(\ref{npcr}) 
does not hold.

The matrix elements of $[{\bf N},{\bf \phi}]$ taken now in the phase basis
$|\phi\rangle$,
\begin{equation}
\langle\phi|[{\bf N},{\bf \phi}]|\phi'\rangle
=(\phi'-\phi)\langle \phi|{\bf N}|\phi'\rangle=i\delta(\phi-\phi'),\;\;\;
-\pi<\phi,\phi'<\pi
\end{equation}
imply that
\begin{equation}
\langle \phi|{\bf N}|\phi'\rangle=-i\frac{\delta(\phi-\phi')}{\phi-\phi'}
=i\frac{d}{d\phi}\delta(\phi-\phi').
\end{equation}
Defining a state $|\psi\rangle=\int_{-\pi}^{\pi}d\phi\;\psi(\phi)|\phi\rangle$
in the basis of the phase states ($\psi(\phi)$ is a complex function of $\phi$), 
we find the expected result
\begin{equation}
\langle\psi|{\bf N}|\psi\rangle=\int_{-\pi}^{\pi}d\phi\int_{-\pi}^{\pi}d\phi'\;
\psi^{*}(\phi)\psi(\phi')\langle \phi|{\bf N}|\phi'\rangle
=\int_{-\pi}^{\pi}d\phi\;\psi^{*}(\phi)\left(i\frac{d}{d\phi}\right)\psi(\phi).
\label{aven}
\end{equation}
Thus, for $\psi(\phi)=\frac{1}{2\pi}\exp(-in\phi)$ we find the desired result
$\langle\psi|{\bf N}|\psi\rangle=n$, but for 
$\psi(\phi)=\frac{1}{2\pi}\exp(+in\phi)$ Eq.~(\ref{aven}) implies that
$\langle\psi|{\bf N}|\psi\rangle=-n$. Furthermore, if  $\psi(\phi)=\psi(-\phi)$
or $\psi(\phi)=-\psi(-\phi)$ then $\langle\psi|{\bf N}|\psi\rangle=0$ which implies
that ${\bf N}|\psi\rangle=0$ since ${\bf N}$ is supposed to be a positive defined
operator. That is, there are many states $|\psi\rangle$ with 
${\bf N}|\psi\rangle =0$. These arguments show why Eq.~(\ref{npcr}) cannot
hold for a number operator which is bounded from below.

Hence, it is the fact that ${\bf N}$ is bounded from below which makes the 
definition of a phase operator problematic. Moreover, if something like 
Eq.~(\ref{npcr}) does not hold, one can raise the question about
the meaning of ``canonical conjugate''. How can we define the
canonical conjugate to ${\bf N}$ without first stating the commutation relation? 

Here we describe a new technique for 
defining the canonical conjugate to {\bf N}. 
The idea is to start from two observables which are canonically conjugate
in the traditional way, 
namely, ${\bf x}$ and ${\bf p}$, and that operate on a Hilbert space which 
is spanned by the eigenstates of ${\bf x}$, or equivalently those of ${\bf p}$ . 
Then, after introducing some conditions
(periodicity, symmetries, etc...), the continuous spectrum of ${\bf p}$ is 
transformed into the spectrum of a number operator. The corresponding
${\bf p}$ and ${\bf x}$ in the new Hilbert space play the role of number ${\bf N}$ and
an absolute value of phase ${\bf\Phi}$ operators, respectively. As we shall see below,
there is no ``simple'' commutation relation (like Eq.~(\ref{npcr})) for 
${\bf N}$ and ${\bf \Phi}$. We shall mention here that our technique looks
similar to the one given by Newton~\cite{Newton}, however, in his approach 
he doubled the Hilbert space of a simple harmonic oscillator. Our method
enables us to define the absolute value of the quantum phase without 
enlarging the physical Hilbert space. 

It is interesting to show first that the requirement of periodicity
alone yields the rotation angle operator of the plane rotator.
We shall see that the angle operator and its canonical conjugate, the angular 
momentum, satisfy Eq.~(\ref{npcr}).
That is, the problem of a free particle in a box (with periodic surface conditions) 
is equivalent to the problem of a plane rotator. 
This motivates us to add the symmetry of reflection and so obtain the canonical
conjugate to the number operator. 

We start with the definition of the Hilbert space ${\cal H}$
of a particle in a free one dimensional space. That is, ${\cal H}$ consist
of all the square integrable functions $u(x)\equiv\langle x|u\rangle$,
with $-\infty<x<\infty$. In the $x$-representation of ${\cal H}$, the
inner product between two elements $u(x),v(x)\in {\cal H}$ is given by 
\begin{equation}
\langle u|v\rangle=\int_{-\infty}^{\infty}u^{*}(x)v(x)dx.
\end{equation}   
The position observable ${\bf x}$ is represented by a multiplication by $x$,
and its conjugate momentum ${\bf p}=i\hbar d/dx$ (since both ${\bf x}$
and ${\bf p}$ are unbounded operators, there are functions lying outside 
their domains of definition). The monochromatic wave   
\begin{equation}
u_{p}(x)\equiv\langle x|p\rangle=(2\pi)^{-1/2}\exp\left(-ixp/\hbar\right)
\label{xp1}
\end{equation}
do not belong to ${\cal H}$, but it satisfies ${\bf p}u_{p}(x)=pu_{p}(x)$.
As we shall see in the following, when the domain of $x$ is restricted,
the momentum eigen-functions do belong to the Hilbert space.
Note that $[{\bf p},{\bf x}]=i\hbar$. We do not use the conventional
$[{\bf p},{\bf x}]=-i\hbar$ due to the analogy to be shown later with 
${\bf N}$ and ${\bf \phi}$. 

Consider now the Hilbert space ${\cal H}_{L}$ 
defined by the condition that 
$\psi(x)\in~{\cal H}_{L}$ if
$\psi (x-L/2) =\psi (x+L/2)$. 
This periodicity condition implies that
the uncertainty in the location of the particle is infinity. Hence, 
instead of working with $-\infty<~x~<+\infty$, we shall restrict 
the domain of $x$ by defining 
$x_{new}=x$~mod~$L$ so that  
$-L/2<~x_{new}~\leq +L/2$. Therefore, the ``new'' position observable 
${\bf x}_{L}$, which 
operates in ${\cal H}_{L}$, is defined by
\begin{equation}
{\bf x}_{L}\;\psi(x)=(x\;{\rm mod}\;L)\psi(x)\;\;\;\forall
\;\psi(x)\in {\cal H}_{L}.
\end{equation}
Note that the uncertainty in the location of the particle is now finite.
If $u_{p}(x)\equiv\langle x|p\rangle\in {\cal H}_{L}$, then 
$\langle x-L/2|p\rangle=\langle x+L/2|p\rangle$ and
according to Eq.~(\ref{xp1}) we find that $p$ must be of 
the form $p_{n}\equiv 2\pi\hbar n/L$ where 
$n=0,\pm 1,\pm 2,...$. It is easy to show that 
any function $\psi(x)\in {\cal H}_{L}$ is a superposition of the
functions $u_{n}(x)\equiv\langle x|p_{n}\rangle$, i.e.
\begin{equation}
\psi(x)=\sum_{n=-\infty}^{\infty}c_{n}u_{n}(x),
\end{equation}
where $c_{n}$ are complex numbers. Accordingly, the momentum observable
which operates in ${\cal H}_{L}$ is defined by
\begin{equation}
{\bf p}_{L}\psi(x)=\sum_{n=-\infty}^{\infty}p_{n}c_{n}u_{n}(x)\;\;\;\forall\;
\psi(x)\in {\cal H}_{L}.
\end{equation}  
 
It is interesting that even though, in analogy with Eq.(\ref{xp1}), 
\begin{equation}
u_{n}(x)\equiv\langle x|p_{n}\rangle=L^{-1/2}\exp\left(-ixp_{n}/\hbar\right),
\label{ccon}
\end{equation}
the familiar commutation relation $[{\bf p},{\bf x}]= i\hbar$ is here
replaced by 
\begin{equation}
[{\bf p}_{L},{\bf x}_{L}] 
=i\hbar\left(1-L\delta\left({\bf x}-\frac{L}{2}\right)\right).
\label{commu}
\end{equation}
Eq.~(\ref{commu}) follows from the matrix elements of $[{\bf p}_{L},{\bf x}_{L}]$
in the momentum basis $u_{n}(x)$:
\begin{eqnarray}
\langle p_{n}|[{\bf p}_{L},{\bf x}_{L}]|p_{n'}\rangle & \equiv &
\int_{-\frac{L}{2}}^{\frac{L}{2}}dx u_{n}^{*}(x)[{\bf p}_{L},{\bf x}_{L}]
u_{n'}(x)=\frac{1}{L}(p_{n}-p_{n'})\int_{-\frac{L}{2}}^{\frac{L}{2}}xe^{\frac{i}
{\hbar}(p_{n}-p_{n'})x}dx\nonumber\\
& = & i\hbar\left[\delta_{n,n'}-e^{-\frac{iL}{2\hbar}
(p_{n}-p_{n'})}\right].
\end{eqnarray}

The commutation relation between the dimensionless operators 
${\bf \Theta}\equiv\frac{2\pi}{L}{\bf x}_{L}$ and 
${\bf\cal N}\equiv\frac{L}{2\pi\hbar}{\bf p}_{L}$
is thus exactly the same as Eq.~(\ref{npcr}). 

Note that ${\bf\cal N}$ can have negative eigenvalues. 
Thus, ${\bf J}_{z}\equiv\hbar{\bf\cal N}$ and ${\bf \Theta}$ can be interpreted
as the angular momentum and the rotation angle of a plane rotator, respectively.
The matrix elements of ${\bf \Theta}$ in the momentum state basis 
$u_{n}(x)\equiv\langle x|n\rangle\equiv\langle x|p_{n}\rangle$ are given by  
\begin{equation}
\langle n|{\bf \Theta}|n'\rangle=\frac{2\pi}{L}\int_{-\frac{L}{2}}^{\frac{L}{2}}
x\langle p_{n}|x\rangle\langle x|p_{n'}\rangle dx
=\frac{i(-1)^{n-n'}}{n-n'}(1-\delta_{nn'}),
\label{matp}
\end{equation}
where we have used Eq.~(\ref{ccon}).
The same matrix elements have been obtained by Galindo~\cite{Galindo} 
and can be obtained also from an earlier work by Garrison and Wong~\cite{Garrison}.
Garrison-Wong-Galindo (GWG) theory claims that the angle operator ${\bf \Theta}$
given by Eq.~(\ref{matp}) is canonically conjugate to the {\it number}
operator and not to the angular momentum ${\bf J}_{z}$. However, GWG 
theory does not pass the Barnnet and Pegg ``acid-test'' because they perform 
calculations with only positive eigenvalues of ${\bf J}_{z}$. In our method, no 
such problem arises since  Eq.~(\ref{ccon}) implies that the angle distribution 
in a momentum (number) state is uniform: 
$P_{n}(\theta)=|\langle n|\theta\rangle|^{2}=1/2\pi$, 
where $|n\rangle\equiv|p_{n}\rangle$ and 
$|\theta\rangle\equiv\sqrt{\frac{L}{2\pi}}|x=\frac{L}{2\pi}\theta\rangle$.
 
The next step in our procedure is to add a condition which will define a new
subspace of ${\cal H}_{L}$ in which the momentum operator has
only positive eigenvalues! However, in order to make the problem clearer,
we shall first find the relevant subspace of the original Hilbert space
${\cal H}$ (i.e. a subspace with {\it continuous} positive momentum).    
In the same way as we made ${\bf x}$ bounded by demanding periodicity we shall
make here ${\bf p}$ positive by demanding symmetry upon reflection in the
$p$-representation.
As we shall see in the following, this requirement leads to a symmetry 
in the $x$-representation as well. 

The subspace ${\cal H}^{+}$ is defined as follows: for all
$\psi(p)\equiv\langle p|\psi\rangle\in {\cal H}^{+}\subset {\cal H}$, 
$\langle p|\psi\rangle =\langle -p|\psi\rangle $. That is,
all states in ${\cal H}^{+}$ are even in the momentum representation.
Furthermore, any such state can be written as 
$\psi(p)\equiv\langle p|\psi\rangle=\int_{-\infty}^{\infty}dx\;u_{p}^{*}(x)\psi(x)$,
where $\psi(x)\equiv\langle x|\psi\rangle$. 
Thus,
Eq.~(\ref{xp1}) and the requirement $\langle p|\psi\rangle =\langle -p|\psi\rangle $
imply that
\begin{equation}
\int_{-\infty}^{\infty}\psi(x)e^{ipx}dx
=\int_{-\infty}^{\infty}\psi(x)e^{-ipx}dx
\end{equation}
for all $p$. Therefore, for $\psi(x)\equiv\langle x|\psi\rangle\in{\cal H}^{+}$,
$\psi(x)$ is even in $x$ (i.e. $\langle x|\psi\rangle =\langle -x|\psi\rangle $). 

One can easily verify that for any $\psi(x)\in {\cal H}$, 
$\psi^{+}(x)\equiv(\psi(x)+\psi(-x))/2$ (or in the $p$-representation
$\psi^{+}(p)\equiv(\psi(p)+\psi(-p))/2\;$) belongs to ${\cal H}^{+}$.
Thus, we shall denote $\psi^{+}(x)\equiv {}^{+}\langle x|\psi\rangle$
(and $\psi^{+}(p)\equiv {}^{+}\langle p|\psi\rangle$) where 
$|x\rangle ^{+}\equiv\frac{1}{2}(|x\rangle +|-x\rangle )$
(and $|p\rangle ^{+}\equiv\frac{1}{2}(|p\rangle +|-p\rangle )$).
Using this notation
\begin{equation}
^{+}\langle x|x^{\prime}\rangle^{+}=\frac{1}{2}\left(\delta(x-x^{\prime})
+\delta(x+x^{\prime})\right)\;\;\;{\rm and}\;\;\; ^{+}\langle p|p^{\prime}\rangle^{+}
=\frac{1}{2}\left(\delta(p-p^{\prime})+\delta(p+p^{\prime})\right).
\end{equation}
The analog function to the monochromatic wave in Eq.~(\ref{xp1}) 
is given by
\begin{equation}
u_{p}^{+}(x)\equiv ^{+}\langle x|p\rangle^{+}=
(2\pi)^{-1/2}\cos\left(xp/\hbar\right).
\label{xppl}
\end{equation}
Note that $u_{p}^{+}(x)$ is not in ${\cal H}^{+}$ because it is not a square 
integrable function. 

Now, since all the functions in ${\cal H}^{+}$ are even,
we can restrict our Hilbert space to the domain $p\geq 0$ and $x\geq 0$.
This is exactly the same idea as restricting the domain of $x$ from 
$(-\infty,\infty)$ to $(-L/2,+L/2]$ in ${\cal H}_{L}$.
We now have to change the 
normalization condition in the following way,
\begin{equation}
|x\rangle^{+}\longrightarrow\sqrt{2}|x\rangle^{+}
\;\;\;{\rm and}\;\;\;|p\rangle^{+}\longrightarrow\sqrt{2}|p\rangle^{+}
\end{equation}
for $x$ and $p$ greater then zero. For $|x=0\rangle^{+}$ and 
$|p=0\rangle^{+}$ no change is needed. Hence,
\begin{equation}
 ^{+}\langle x|x^{\prime}\rangle^{+}\equiv 
\int_{0}^{\infty}dp\;u_{p}^{*}(x)u_{p}(x')
= \delta(x-x^{\prime})
\;\;\;{\rm for}\;\;\;x,x^{\prime}>0\;\;\;{\rm and}\;\;\;
^{+}\langle x|x^{\prime}=0\rangle^{+}=2\delta(x)
 \label{norm}
\end{equation}
and analogous conditions for $p$. 
The function in Eq.~(\ref{xppl}), normalized to the domain $p\geq 0$ and $x\geq 0$,
is given by
\begin{equation}
u_{p}^{+}(x)=
\sqrt{\frac{2}{\pi}}\cos\left(xp/\hbar\right).
\label{xpp2}
\end{equation}
The {\it positive} 
position operator is defined as follows: ${\bf x}^{+}\psi(x)=|x|\psi(x)$
for all $\psi(x)\in {\cal H}^{+}$. Now, since any $\psi(x)\in {\cal H}^{+}$
can be written as $\psi(x)=\int_{0}^{\infty}dp\;c_{p}u_{p}(x)$ ($c_{p}$ are
complex numbers), the positive momentum observable can be defined as follows:
\begin{equation}
{\bf p}^{+}\psi(x)=\int_{0}^{\infty}dp\;p\;c_{p}u_{p}(x)\;\;\;\forall\;
\psi(x)\in {\cal H}^{+}.
\end{equation}
 
One can also verify that
$[{\bf x}^{+},{\bf p}^{+}]\ne i\hbar$; it is actually impossible
to express the commutator in such a simple form. Nevertheless, 
${\bf x}^{+}$ and ${\bf p}^{+}$ are observables which are canonically 
conjugate. They represent the momentum and the position of
the particle in the half-spaces. It should be clear that in the process of ``making''
${\bf p}$ positive, its conjugate position ${\bf x}$ turns out to be positive
too! It is impossible to cut half of the momentum space
without changing the position space. The idea that in half space $x\geq 0$
the momentum is positive, means that free particles move towards 
$x\rightarrow\infty$ but not towards $x=0$ (because $x<0$ is not defined). 
As we shall see, this means that
since the number operator is positive, its canonical conjugate will be also 
positive, i.e. it prevents the existence of a phase operator ${\bf\phi}$,
but not of its absolute value ${\bf\Phi}$.

In order to obtain an absolute phase operator we define a new Hilbert space, 
${\cal H}^{+}_{L}\subset {\cal H}_{L}$ as follows: If 
$\psi(x)\in{\cal H}^{+}_{L}\subset {\cal H}_{L}$ then $\psi (x)=\psi (-x)$. 
In this Hilbert space the momentum operator 
is positive, discrete and with uniform intervals between eigenvalues. It is 
defined by its eigen-functions $u_{n}(x)$:
\begin{equation}
u_{n}(x)\equiv {}^{+}\langle x|p_{n}\rangle^{+}=\sqrt{4/L}\cos
\left(xp_{n}/\hbar\right)\;\;\;{\rm for}\;\;\;n\ne 0
\;\;\;{\rm and}\;\;\;^{+}\langle x|p_{n=0}\rangle^{+}=\sqrt{2/L},
\label{xpre}
\end{equation}
where we have used Eq.~(\ref{xpp2}) and $p_{n}=2\pi\hbar n/L\;(n=0,1,2...)$.
Hence, the positive momentum observable ${\bf p}_{L}^{+}$ is defined by:
${\bf p}_{L}^{+}\;u_{n}(x)=p_{n}u_{n}(x)$.
The conjugate position operator ${\bf x}^{+}_{L}$ is also positive and has 
eigenvalues $0\leq x\leq L/2$: 
\begin{equation}
{\bf x}^{+}_{L}\psi(x)=|x\;{\rm mod}\;L|\psi(x)\;\;\;\forall\;
\psi(x)\in {\cal H}_{L}^{+}. 
\end{equation}  

Now, the number and the absolute phase operators can be defined by
\begin{equation}
{\bf N}\equiv\frac{L}{2\pi\hbar}{\bf p}^{+}_{L}\;\;\;
{\rm and}\;\;\;{\bf\Phi}\equiv\frac{2\pi}{L}{\bf x}^{+}_{L},
\label{defp}
\end{equation}
where the number eigen-functions are given by
\begin{equation}
u_{n}(\Phi)\equiv\langle\Phi|n\rangle\equiv\sqrt{\frac{L}{2\pi}}
{}^{\;+}\langle x=\frac{L}{2\pi}\Phi|p_{n}\rangle ^{+}=
\left\{
\begin{array}{ll}
& \sqrt{2/\pi}\cos(n\Phi)\;\;{\rm if}\;\;n>0\\
& 1/\sqrt{\pi}\;\;{\rm if}\;\;n=0
\end{array}
\right.
\label{gili}
\end{equation} 
 Using Eq.~(\ref{gili}) we find that
\begin{equation}
|\Phi\rangle=\frac{1}{\sqrt{\pi}}|n=0\rangle+\sqrt{\frac{2}{\pi}}
\sum_{n=1}^{\infty}\cos(n\Phi)|n\rangle.
\label{phin}
\end{equation}
It can be shown that the above equation is consistent with the normalization 
conditions $\langle\Phi|\Phi'\rangle=\delta(\Phi-\Phi')$ for $0<\Phi,\Phi'\leq\pi$
and $\langle\Phi|\Phi=0\rangle=2\delta(\Phi)$.

Eq.~(\ref{phin}) implies that the distribution of the absolute phase in 
a number state $|n\rangle$ is given by
\begin{eqnarray}
P_{n=0}(\Phi) & \equiv & |\langle n=0|\Phi\rangle|^{2}=\frac{1}{\pi}\nonumber\\
P_{n>0}(\Phi) & \equiv & |\langle n|\Phi\rangle|^{2}=\frac{2}{\pi}\cos^{2}(n\Phi)= 
\frac{1}{\pi}(1+\cos(2n\Phi)).
\label{nln}
\end{eqnarray}
It is not surprising that the distribution is not uniform since it is the 
distribution of the absolute value of the phase, and not of the phase itself.
In the case of the plane rotator for example, the state 
$|\theta\rangle_{+}\equiv 1/\sqrt{2}(|\theta\rangle+|-\theta\rangle)$ is an 
eigenstate of the absolute value of the angle operator, and it gives a similar 
distribution as in Eq.~(\ref{nln}).  
Thus, Eq.~(\ref{nln}) yields
a non-uniform phase distribution. However, in the classical limit 
$n\rightarrow\infty$, the average of $\Phi^{m}$ ($m=0,1,2...$) in a number 
state is given by:
\begin{equation}
\langle n| {\bf\Phi}^{m}|n\rangle\equiv\langle\Phi^{m}\rangle_{n}
=\frac{1}{\pi}\int_{0}^{\pi}d\Phi(1+\cos(2n\Phi))\Phi^{m}
\;\rightarrow\;\frac{1}{\pi}\int_{0}^{\pi}d\Phi\Phi^{m}
\label{avnu}
\end{equation}
which is identical to the average
of $\Phi^{m}$ in a classical uniform phase distribution $P(\Phi)=1/\pi$.

The matrix elements of ${\bf\Phi}$ in the $|n\rangle$ basis are determined
from its definition in Eq.~(\ref{defp}) and from Eq.~(\ref{phin})
\begin{equation}
\langle n|{\bf\Phi}|n'\rangle=\langle n'|{\bf\Phi}|n\rangle=
\left\{
\begin{array}{ll}
& \pi/2\;\;\;\;{\rm for}\;\;n'=n\\
& -2\sqrt{2}\pi^{-1}n^{-2}
\;\;\;\;{\rm for}\;\;n'=0,\;\; {\rm odd}\;\;n\\
& -2\pi^{-1}
\left((n+n')^{-2}+(n-n')^{-2}\right)\;\;\;{\rm for}
\;\;\;n,n'>0,\;\;{\rm odd}\;\; n+n'\\
& 0\;\;\;{\rm otherwise}.
\end{array}
\label{gg}
\right.
\end{equation}
By comparing our method with that of Carruthers and Nieto~\cite{CandN},
we find a correction to the ``phase cosine'' 
${\bf C}\equiv\frac{1}{2}({\bf E}+{\bf E}^{\dag})$
and ``phase sine'' ${\bf S}\equiv\frac{1}{2i}({\bf E}-{\bf E}^{\dag})$
\begin{eqnarray}
\cos{\bf\Phi} & = & \int_{0}^{\pi}\cos\Phi\;|\Phi\rangle\langle\Phi|\;d\Phi
={\bf C}+\frac{1}{2}(\sqrt{2}-1)(|0\rangle\langle 1|+|1\rangle\langle 0|)\nonumber\\
\sin^{2}{\bf\Phi} & = & \int_{0}^{\pi}\sin^{2}\Phi\;|\Phi\rangle\langle\Phi|\;d\Phi
={\bf S}^{2}+\frac{1}{4}(1-\sqrt{2})(|0\rangle\langle 2|+|2\rangle\langle 0|)
+\frac{1}{4}(|0\rangle\langle 0|-|1\rangle\langle 1|),
\label{cands}
\end{eqnarray}
where the projectors involving number eigenstates 
$|0\rangle,|1\rangle,|2\rangle$ can be neglected 
for states with $\langle {\bf N}\rangle\gg 1$. Note, 
that we have used here the Dirac notation for operators.

It is interesting to calculate the expectation values of functions of 
${\bf \Phi}$ in a coherent state 
\begin{equation}
\langle\Phi|\gamma\rangle\equiv\exp(-\frac{1}{2}|\gamma|^{2})
\sum_{n=0}^{\infty}\frac{\gamma^{n}}{\sqrt{n!}}u_{n}(\Phi), 
\end{equation}
where $\gamma=\sqrt{N}e^{i\theta}$ is the eigenvalue of the annihilation
operator ${\bf a}$.
It can be shown~\cite{Gour} that in the classical limit 
$N=|\gamma|^{2}\rightarrow\infty$,
\begin{eqnarray}
\langle\gamma|{\bf\Phi}|\gamma\rangle & \rightarrow & \frac{\pi}{2}
-\frac{4}{\pi}\sum_{s=1,3,5...}\frac{\cos(\theta s)}{s^{2}} 
=|\theta|\nonumber\\
\langle\gamma|\sin({\bf\Phi})|\gamma\rangle & \rightarrow & \frac{2}{\pi}
-\frac{4}{\pi}\sum_{s=2,4,6...}\frac{\cos(\theta s)}{s^{2}-1}
=|\sin\theta|\nonumber\\
\langle\gamma|\cos({\bf\Phi})|\gamma\rangle & \rightarrow & \cos\theta\nonumber\\
\langle\gamma|\cos^{2}({\bf\Phi})|\gamma\rangle & \rightarrow & 
\cos^{2}\theta\nonumber\\
\langle\gamma|\sin^{2}({\bf\Phi})|\gamma\rangle & \rightarrow & 
\sin^{2}\theta
\label{clim}
\end{eqnarray}
where the last three limits can be obtained from Eq.~(\ref{cands}).
These results prove useful for establishing that ${\bf\Phi}$ has 
the correct large-field correspondence limit. The first two limits 
in Eq.~(\ref{clim}) again show that the operator ${\bf\Phi}$ corresponds to the 
{\it absolute} value of the phase observable.

It was suggested by Pegg and Barnett~\cite{Pegg} to calculate all
physical results in a finite dimensional space and only then to take the limit of 
the dimension to infinity. In our model, a finite dimensional space can be obtained
by adding to ${\cal H}_{L}$ the requirement of periodicity in the momentum space.
In such a Hilbert space the momentum is restricted by $-m\leq n\leq m$ or,
{\it equivalently} $0\leq n\leq 2m$. However, the limit $m\rightarrow\infty$
can be carried out only in the case where $-m\leq n\leq m$ and {\it not}
in the case where $0\leq n\leq 2m$ (this will be shown in future work). Hence,
in order to obtain positive $n$ we must require symmetry upon reflection, and
not periodicity in the momentum space.

\section*{Acknowledgments}
I would like to thank J.~Bekenstein 
for his guidance and support. This research was supported by grant 
No. 129/00-1 of the Israel Science Foundation,
and by a Clore Foundation fellowship.

\end{document}